\documentclass[conference]{IEEEtran}
\pdfoutput=1
\IEEEoverridecommandlockouts
\usepackage{amsmath,amssymb,amsfonts}
\usepackage{algorithmic}
\usepackage{graphicx}
\usepackage{booktabs}
\usepackage{array}
\usepackage{tabularx}
\usepackage{float}
\usepackage{textcomp}
\usepackage{xcolor}
\usepackage{makecell}

\usepackage{amsmath}
\usepackage{tabularx}
\usepackage{multirow}
\usepackage{makecell}

\usepackage[skip=5pt]{caption}
\usepackage{caption}
\RequirePackage[backend=biber,
natbib=true,
style=numeric-comp,
url=false,
doi=false,
sorting=none]{biblatex}

\def\BibTeX{{\rm B\kern-.05em{\sc i\kern-.025em b}\kern-.08em
    T\kern-.1667em\lower.7ex\hbox{E}\kern-.125emX}}
\begin{document}

\title{Low Complexity Joint Chromatic Dispersion and Time/Frequency Offset Estimation Based on Fractional Fourier Transform\\

}

\author{
    \IEEEauthorblockN{Guozhi Xu, Zekun Niu, Lyu Li, Weisheng Hu, Lilin Yi*}
    \IEEEauthorblockA{
        \textit{State Key Lab of Advanced Optical Communication Systems and Networks,}\\
        \textit{School of Electronic Information and Electrical Engineering,}\\
        \textit{Shanghai Jiao Tong University,}\\
        Shanghai, China\\
        e-mail: lilinyi@sjtu.edu.cn
    }
}


\maketitle

\begin{abstract}
We propose and experimentally validate a joint estimation method for chromatic dispersion and time-frequency offset based on the fractional Fourier transform, which reduces computational complexity by more than 50\% while keeping estimation accuracy.
\end{abstract}
\renewcommand\IEEEkeywordsname{Keywords}

\section{Introduction}
For high speed long distance coherent optical communication, the accuracy of linear parameters such as chromatic dispersion (CD), time offset (TO), and frequency offset (FO) directly impacts system performance. To ensure effective signal recovery, it is crucial to perform accurate and efficient estimation of these parameters before applying compensation in the receiver digital signal processing (DSP) unit. Numerous estimation methods have been proposed, with training sequence (TS)-based methods standing out for their modulation format independence and high accuracy[1,2]. However, these methods for estimating different parameters are usually executed independently, leading to high computational complexity and redundant TSs. 

The fractional Fourier transform (FrFT), an extension of the Fourier transform (FT), has emerged as a powerful tool for addressing time-frequency problems in communication systems. In recent years, several estimation methods based on FrFT have been proposed. However, the methods discussed in [3-5] focus solely on TO and FO, neglecting CD estimation. Additionally, the methods in [6,7] require the insertion of multiple TSs at the transmitter (Tx) to separately estimate CD, TO and FO, which require multiple estimations at the receiver (Rx) to determine each parameter individually, leading to high time slot usage and significant computational complexity.

In this paper, we propose a low-complexity joint estimation method for CD and time-frequency offset based on FrFT. Compared to the traditional method, our method reduces computational complexity by more than 50\% while maintaining estimation performance and is also applicable to non-integer sampling rate. The universality and robustness of the proposed method have been verified by simulations and experiments for 60 GBaud DP-16QAM 21-channel optical transmission system. The simulation results indicate that the method demonstrates an average chromatic dispersion estimation (CDE) errors of less than 50 ps/nm for 1.25 times sampling rate at the Rx, supports a frequency offset estimation (FOE) range of up to $\pm$8 GHz, and maintains time offset estimation (TOE) errors of less than 1.6 symbol. The long-distance experimental results further confirm the effectiveness of the proposed method.

\section{Principle}
\subsection{Principle of Estimation}
FrFT induces order-dependent rotations in time-frequency plane (Wigner plane). When a direct current (DC) signal is transformed using a P-order FrFT, a chirp signal is produced with a linearly changing instantaneous frequency over time[3]. We utilize the TS at the Tx, which is composed of two chirp signals with opposite orders:
\begin{equation}
TS(t) = \mathcal{F}^{\alpha} [S(t)] + \mathcal{F}^{-\alpha} [S(t)],
\end{equation}
where \( S(t) \) is a DC signal, \( \mathcal{F}^{\alpha} \) denotes the FrFT transformation with a rotation angle \(\alpha = P \cdot \pi / 2 \) and \( TS(t) \) stands for the TS. In the absence of external factors, TS achieves impulse peaks at two optimal rotation angles: 
\(\phi_1 = (1 - P) \cdot \pi / 2 \), \(\phi_2 = -(1 - P) \cdot \pi / 2\).

\begin{figure}[t]
    \centering
    \includegraphics[width=\columnwidth]{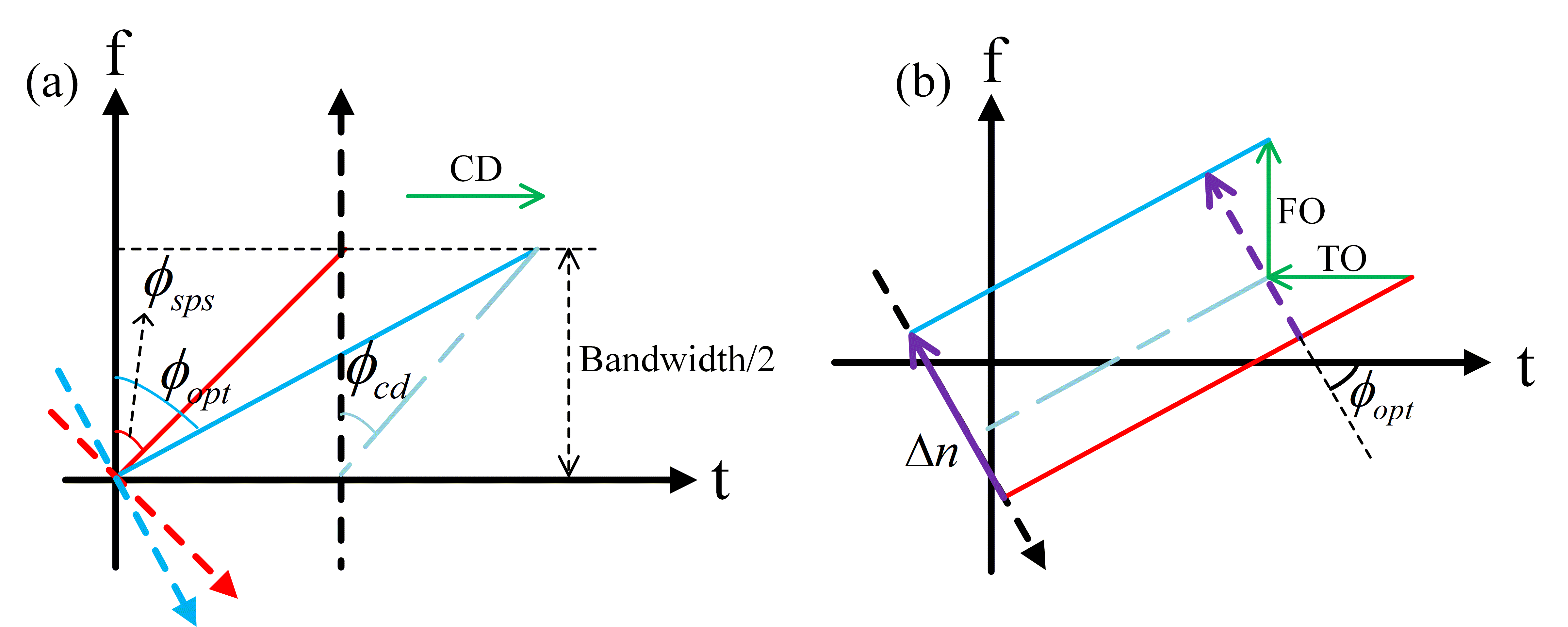}
    \caption{ Time-frequency distribution of chirp signal which is influenced by (a) sps and CD, (b) TO and FO.}
    \label{fig:1}
\end{figure}

However, the optimal rotation angles of TS change at the Rx due to the influence of CD and sampling rate. To simplify our analysis, we consider only one chirp signal, with its expression and corresponding optimal rotation angle as follows:
\begin{equation}
\begin{gathered}
f[n] = \exp\left(j \pi \rho (nT)^2 \right), \quad 0 \leq n \leq N_s - 1, \\
\phi = -\arctan\left(\frac{1}{\rho N_s T^2}\right),
\end{gathered}
\end{equation}
where $\rho$ is the chirp rate and $T$ is the symbol period. After taking into account the Rx sampling rate, the optimal rotation angle is determined to be:
\begin{equation}
\phi_{\text{\textit{sps}}} = -\arctan\left(\frac{\text{\textit{sps}}}{\rho N_s T^2}\right),
\end{equation}
where \(\textit{sps}\) is the sampling rate at the Rx. Therefore, $\tan \phi_{\textit{sps}} = \textit{sps} \cdot \tan \phi$. Additionally, we consider CD as chirp, the corresponding optimal rotation angle is:
\begin{equation}
\phi_{\text{\textit{cd}}} = -\arctan\left(\frac{\text{\textit{sps}} \cdot 2\pi\beta_2 z}{N_s T^2}\right),
\end{equation}
where $\beta_2$ is the group velocity dispersion parameter and $z$ is the transmission distance. As depicted in Fig.~\ref{fig:1}(a), due to the constant bandwidth:
\begin{equation}
\tan \phi_{\text{\textit{opt}}} = \tan \phi_{\text{\textit{sps}}} + \tan \phi_{\text{\textit{cd}}},
\end{equation}
where $\phi_{\text{\textit{opt}}}$ is the optimal rotation angle at the Rx.

After obtaining the optimal rotation angle, we analyze the peak shift in the time-frequency domain. In the Wigner plane, TO and FO shift the peak along the time and frequency axes respectively. As depicted in Fig.~\ref{fig:1}(b),  according to the geometrical analysis:
\begin{equation}
\Delta n = \Delta t \cos \phi_{\text{\textit{opt}}} + \Delta f \sin \phi_{\text{\textit{opt}}},
\end{equation}
where \( \Delta n \) is the center peak offset of TS, \( \Delta t \) and \( \Delta f \) represent the digital sample points number for TO and FO respectively. Since the TS comprises two chirp signals, we can solve a system of simultaneous equations to determine \( \Delta t \) and \( \Delta f \).
\subsection{Process of Estimation}

\begin{figure}[t]
    \centering
    \includegraphics[width=\columnwidth]{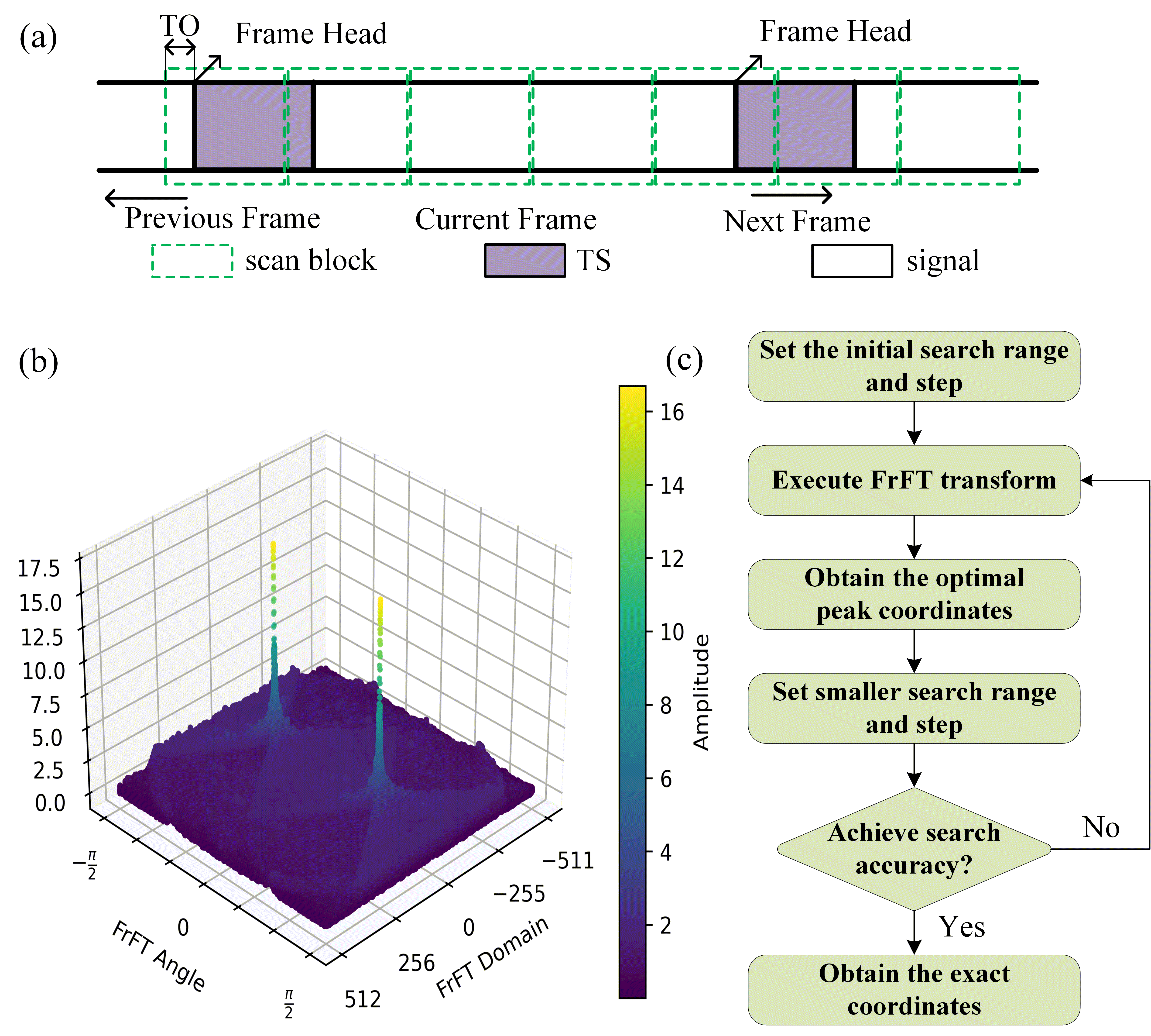}
    \caption{(a) Coarse framing. (b) FrFT Domain-Angle distribution of TS with $\alpha=\pi / 4$ and $N_s=1024$. (c) Optimal coordinates search flow chart.}
    \label{fig:2}
\end{figure}

We generate the TS using (1) and insert it at the beginning of each frame at the Tx. Therefore, the first task at the Rx is TS localization. The method in [6] becomes ineffective for TS localization due to the unknown optimal rotation angles. Here we design a new detection parameter:
\begin{equation}
M[k] = \sum \mathcal|{F}^{\phi_1} [R_k(t)]|,
\end{equation}
where \( R_k (t) \) represents the \( k^{\mathit{th}} \) block of the received signals and the block size is \( N_s \cdot \textit{sps} \). Although \( \phi_1 \) is no longer the optimal rotation angle, the amplitude integral of the TS is significantly lower than that of ordinary signals due to residual energy concentration characteristics. By detecting the minimum \( M[k] \) of signal blocks, we can roughly locate the position of TS, as illustrated in Fig.~\ref{fig:2}(a).

Next, we apply the FrFT with a range of rotation angles to the approximately located TS, creating a two-dimensional search space defined by the FrFT domain and its corresponding rotation angles. Fig.~\ref{fig:2}(b) illustrates the two-dimensional distribution of TS with $\alpha=\pi / 4$ and $N_s=1024$. To reduce complexity, we utilize a stepwise search method to determine the peak coordinates $\{\phi_{\text{\textit{opt}}1}, \Delta n_1\}$ and $\{\phi_{\text{\textit{opt}}2}, \Delta n_2\}$. As shown in Fig.~\ref{fig:2}(c), the initial stage employs a wide search range of \( \pm \pi / 2 \) with a step size of \( 0.01\pi \) to provide a rough search, which is followed by a finer search within a smaller range of \( \pm 0.1\pi \) with a step size of \( 0.001\pi \). Generally, two iterations are sufficient to accurately determine the coordinates of the two peaks. If higher precision is required, the iterative process can be continued.

Finally, by substituting the coordinates $\{\phi_{\text{\textit{opt}}1}, \Delta n_1\}$ and $\{\phi_{\text{\textit{opt}}2}, \Delta n_2\}$, along with the known \(\textit{sps}\), into (3), (4), (5) and (6), we can simultaneously determine CD, FO, and TO. Solving the system of equations provides two CD estimates, which are averaged to improve the accuracy of the estimation.
\begin{figure*}[t]
    \centering
    \includegraphics[width=0.9\textwidth]{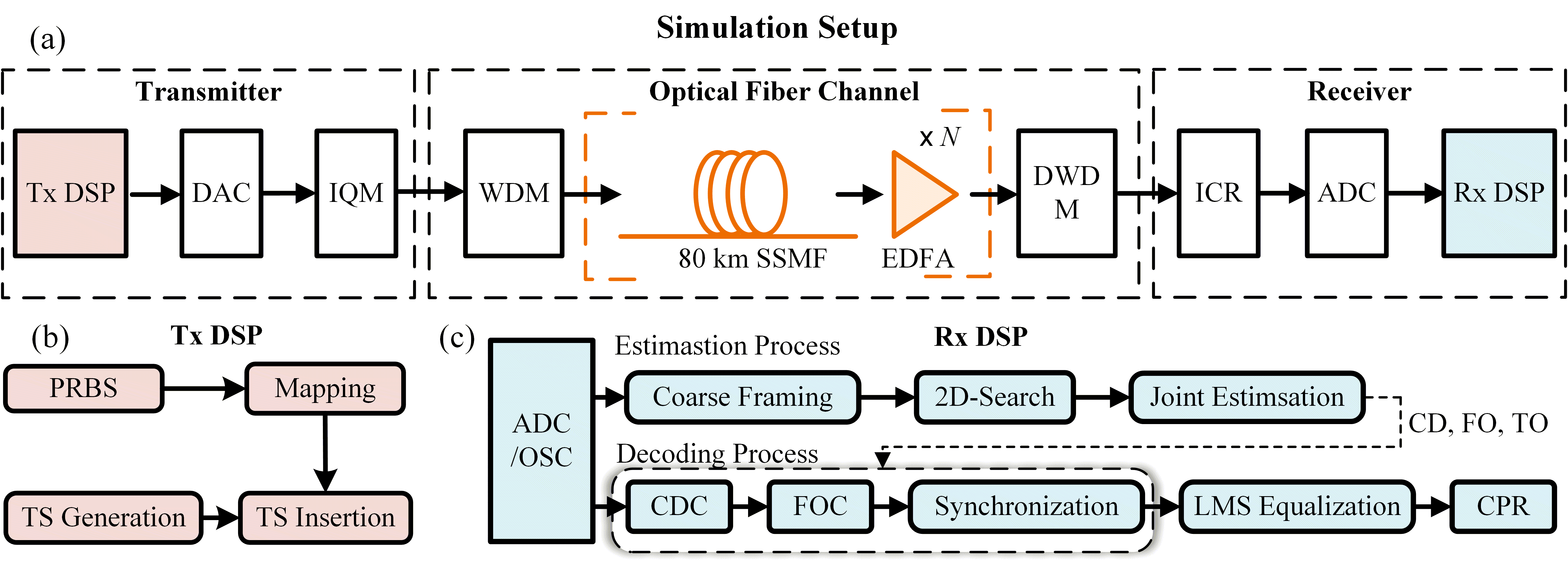}
    \caption{ (a) Simulation setup. (b) Tx DSP. (c) Rx DSP.}
    \label{fig:3}
\end{figure*}
\section{Simulation and Experimental Results}
To verify the universality and robustness of the proposed method, we conducted simulations and experiments for 60 GBaud DP-16QAM 21-channel optical transmission system, using the same Tx and Rx DSP.
\subsection{Simulation Setup and Results}
As shown in Fig.~\ref{fig:3}(a), in the simulation system, signals are processed through the wave division multiplexing (WDM) module, and the input power is adjusted before entering the optical fiber. The signals are then launched into a fiber link composed of 25 spans of standard single-mode fiber (SSMF), each with a span length of 80 km and an average CD parameter of 17 ps/km/nm. After each span, the loss is compensated by erbium-doped fiber amplifiers (EDFAs) with an average noise figure of 5.0 dB. Subsequently, the system demultiplexes WDM and sends the signals to the Rx.

At the Tx, an IQ modulator (IQM) driven by an uncorrelated pseudo-random binary sequence (PRBS) of length $2^{16}$ generates 60 GBaud DP-16QAM symbols shaped by a squared root raised cosine filter with a roll-off factor of 0.1. The TS with $\alpha=0.5$ and $N_s=1024$ is inserted at the head of each frame, as illustrated in Fig.~\ref{fig:3}(b). At the Rx, the signals are sampled by the analog-to-digital (ADC) converter and sent to the Rx DSP. After getting parameters through proposed estimation process, the traditional decoding process can be applied, as shown in Fig.~\ref{fig:3}(c).

\begin{figure*}[t]
    \centering
    \includegraphics[width=0.9\textwidth]{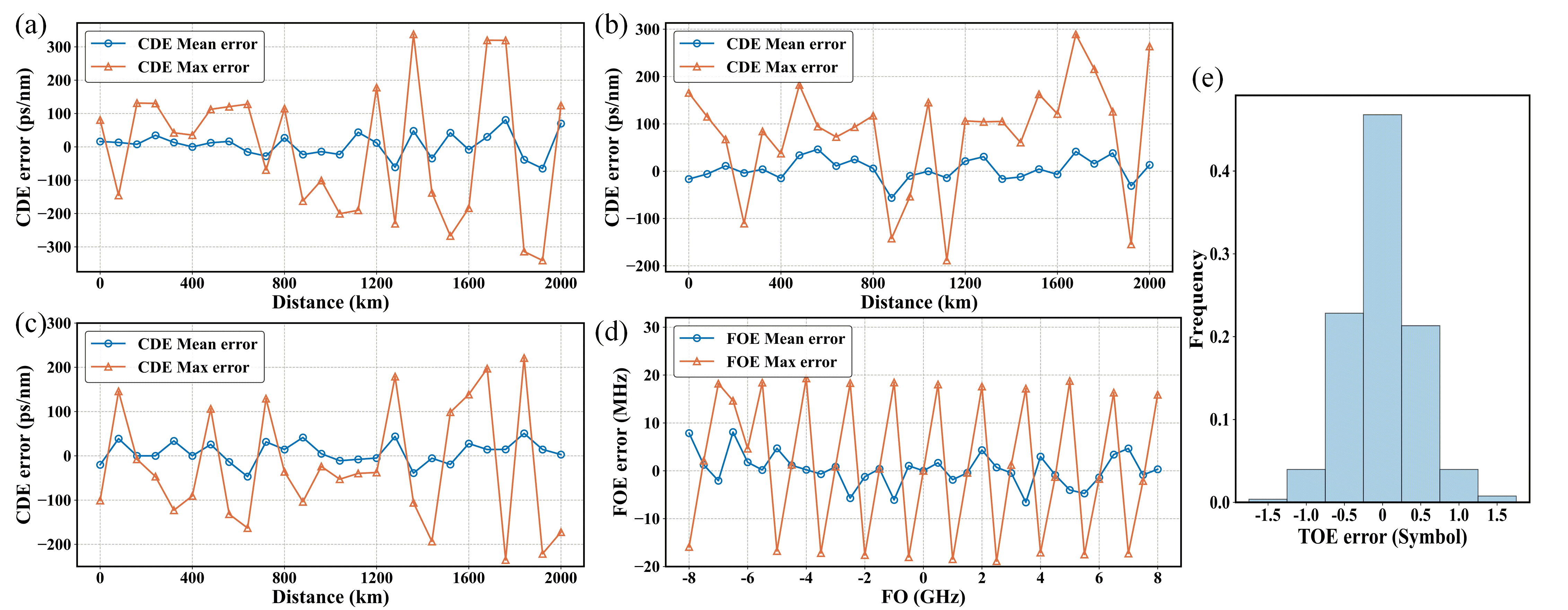}
    \caption{(a) CDE errors for $\textit{sps}=1$. (b) CDE errors for $\textit{sps}=1.25$. (c) CDE errors for $\textit{sps}=2$. (d) FOE errors for $\textit{sps}=1.25$. (e) TOE errors for $\textit{sps}=1.25$.}
    \label{fig:4}
\end{figure*}

To verify the universality of the proposed method, we used the simulation system to test its estimation performance under various conditions. Firstly, we evaluated the CDE performance of our proposed method over various transmission distances for scenarios with $\textit{sps}=1$, $\textit{sps}=1.25$, and $\textit{sps}=2$. The tests were conducted within the range of 0 to 2000 km in increments of 80 km, with 50 groups tested for each distance. The mean estimation errors and the worst-case estimation errors are presented in Fig.~\ref{fig:4}(a), (b), and (c). At $\textit{sps}=2$, the proposed method demonstrates optimal performance, with maximum CDE errors of approximately 200 ps/nm and average errors of less than 50 ps/nm. At $\textit{sps}=1$, the performance degrades, though the estimation accuracy remains within acceptable limits, showing maximum CDE errors of around 300 ps/nm and average errors below 100 ps/nm. The results confirm that our method can produce accurate and effective estimation under CD conditions ranging from 0 to 34,000 ps/nm, even at a undersampling rate.

Additionally, we evaluated the FOE and TOE performances of the proposed method at non-integer sampling rate, with transmission lengths ranging from 0 to 2000 km. For the FOE tests, FO was incremented by 0.5 GHz across a range of -8 GHz to 8 GHz. For the TOE tests, we utilized random TOs at each transmission length. Each condition was tested with 50 sets. The results are illustrated in Fig.~\ref{fig:4}(d)(e) for $\textit{sps}=1.25$. For the FOE tests, the maximum FOE errors are around 20 MHz, with average errors typically below 10 MHz, which are within the tolerance of the subsequent DSP. Additionally, the resolution of digital sampling causes oscillation in the maximum FOE errors, which is consistent with the description in [5]. For the TOE tests, since the TOE performances are consistent across the transmission lengths range, we used frequency distribution histograms to illustrate the results. There is approximately a 50\% frequency of no error, and the maximum TOE error is only 1.6 symbol. While the proposed method may result in a non-zero TOE error, only 10 additional training symbols for an extra correlation operation can ensure perfect timing synchronization.

\begin{figure*}[t]
    \centering
    \includegraphics[width=0.8\textwidth]{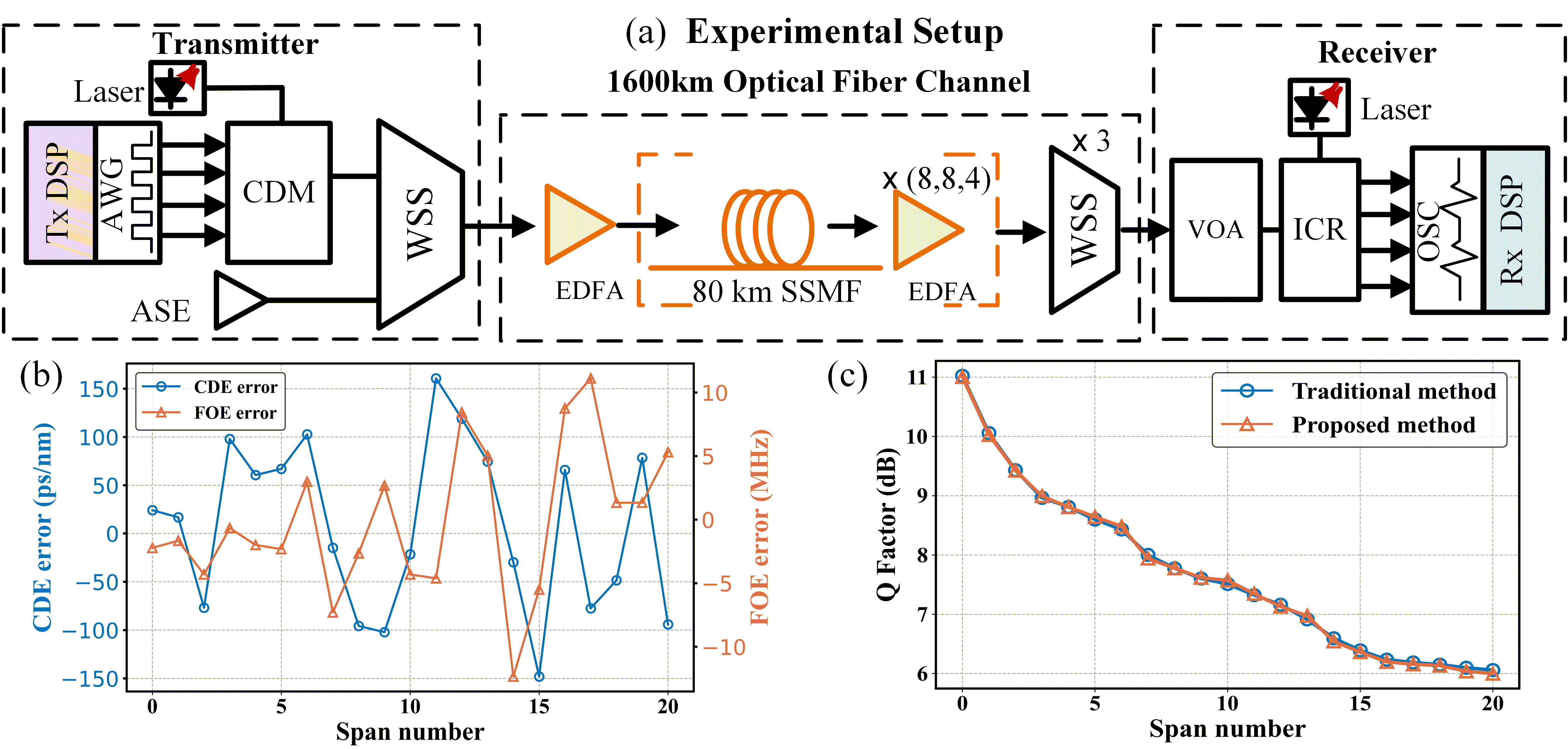}
    \caption{(a) Experimental setup. (b) CDE and FOE errors of the proposed method. (c) Comparison of Q-factor between the two methods.}
    \label{fig:5}
\end{figure*}
\subsection{Experimental Setup and Results}
The experimental setup is similar to the simulation, using the same Tx and Rx DSP, as shown in Fig.~\ref{fig:5}(a). At the Tx, a 120 GSa/s arbitrary waveform generator (AWG) with 40 GHz bandwidth is used to generate 60 GBaud DP-16QAM symbols, using a roll-off factor of 0.1 to shape the signals. The signals are then modulated using a coherent driver modulator (CDM). Wavelength selective switches (WSSs) are placed every 8 spans to maintain a channel space of 75 GHz. After 1600 km of transmission, the signal power is adjusted using a variable optical attenuator (VOA). The signals are then received by an integrated coherent receiver (ICR) with 40 GHz bandwidth and sampled by a 100 GSa/s digital oscilloscope (OSC).

The robustness and actual performance of the proposed method were evaluated using the experimental system. Fig.~\ref{fig:5}(b) illustrates the CDE errors and FOE errors of the proposed method during the experiment. Due to the time-varying nature of FO, we used the 4-power FFT algorithm [9] as the reference standard for the true FO. The experimental results indicate that the maximum CDE error is approximately 150 ps/nm and the maximum FOE error is about 10 MHz, both within the tolerance limits of subsequent DSP.

Finally, We compared the proposed method with the traditional method. In the traditional method, CDE is performed using the method proposed in [8], which is also based on FrFT. FOE utilizes the 4-power FFT algorithm, and TOE is achieved via the cross-correlation operation. Fig.~\ref{fig:5}(c) presents the Q-factor for both methods in the experimental system. The results indicate that both methods exhibit similar transmission performance, which proves the effectiveness of our proposed method.


\subsection{Complexity Analysis}
\newcolumntype{C}{>{\centering\arraybackslash}m{3cm}} 
\newcolumntype{Y}{>{\centering\arraybackslash}X}
\begin{table}[htbp]
\centering
\caption{Estimation Process Comparison}
\begin{tabularx}{\columnwidth}{|C|Y|Y|}
\hline
\textbf{Estimation Process} & \textbf{The Proposed Method} & \textbf{The Traditional Method} \\
\hline
Coarse Framing & $M \log N$ & / \\
\hline
Two-Dimensional Search & $KN \log N$ & / \\
\hline
CDE[1] & / & $2KN \log N$ \\
\hline
FOE (4-power FFT) & / & $N \log N$ \\
\hline
TOE (Cross-Correlation) & / & $MN$ \\
\hline
Total & \makecell{$(M + KN) \log N$} & \makecell{$(2KN + 1) \log N$ \\ $+ MN$} \\
\hline
\end{tabularx}
\label{tab1}
\end{table}
\vspace{-10pt}
We analyzed and compared the computational complexity of the proposed method with the traditional method. For simplicity, we focused on the total number of multiplications required, considering only the primary complexity components of each method. Let the frame length be $M$, the TS length be $N$, and the number of chirp scans be $K$. Both N-point FrFT and N-point FFT require $N\log N$ multiplications. As shown in Table.~\ref{tab1}, the proposed method significantly reduced the total number of multiplications compared to the traditional method, resulting in a more than 50\% reduction in overall complexity. Additionally, the traditional method required dispersion compensation between CDE and time-frequency synchronization. In contrast, the proposed method decouples the estimation and decoding processes, making the estimation independent of dispersion compensation, which further enhances the complexity advantage of the proposed method, particularly when accounting for the computational cost associated with dispersion compensation.

\section{Conclusion}
In this paper, we propose a low-complexity method for joint CD, TO and FO estimation based on the FrFT, which is compatible with non-integer sampling rate. The universality and robustness of this method are demonstrated through simulations and experiments for 60 GBaud DP-16QAM 21-channel optical transmission system. It is worth emphasizing that our method decouples the estimation and decoding processes, which provides a fast and accurate estimation scheme for flexible future optical networks.



\section*{Acknowledgment}

The authors acknowledge the funding provided by the National Key R\&D Program of China (2023YFB2905400), National Natural Science Foundation of China (62025503), and Shanghai Jiao Tong University 2030 Initiative. 

\printbibliography{}
\end{document}